\documentclass[12pt]{article}
\usepackage{amsmath,amsfonts,amssymb,graphicx,float}
\usepackage[utf8]{inputenc}
\usepackage[T1]{fontenc}

\usepackage[top=3cm, bottom=3cm, left=2.5cm, right=2.5cm]{geometry}
\numberwithin{equation}{section}

\begin{document}

\pagestyle{empty}
\vspace*{5mm}

\hfill SU-ITP-12/10

\begin{center}
\vspace{1cm}
{\Large\bf  Radion Dynamics and Phenomenology \\in the Linear Dilaton Model}\\

\vspace{1cm}
{\sc Peter Cox$^{a,}$\footnote{Email: pcox@physics.unimelb.edu.au} {\small and}
Tony Gherghetta$^{a,b,}$\footnote{Email:  tgher@unimelb.edu.au}}\\
\vspace{0.5cm}
 {\it \small {$^a$ARC Centre of Excellence for Particle Physics at the Terascale,\\
 School of Physics, University of Melbourne, Victoria 3010,
Australia}}\\
{\it \small {$^b$Stanford Institute of Theoretical Physics, \\Stanford University, Stanford, CA 94305, USA}}
\vspace{.4cm}
\end{center}

\vspace*{\fill}
\begin{abstract}
We investigate the properties of the radion in the 5D linear dilaton model arising from Little String Theory. 
A Goldberger-Wise type mechanism is used to stabilise a large interbrane distance, with the dilaton now playing the role of the stabilising field. We consider the coupled fluctuations of the metric and dilaton fields and identify the physical scalar modes of the system. The wavefunctions and masses of the radion and Kaluza-Klein modes are calculated, giving a radion mass of order the curvature scale. As a result of the direct coupling between the dilaton and Standard Model fields, the radion couples to the SM Lagrangian, in addition to the trace of the energy-momentum tensor. The effect of these additional interaction terms on the radion decay modes is investigated, with a notable increase in the branching fraction to $\gamma\gamma$. We also consider the effects of a non-minimal Higgs coupling to gravity, which introduces a mixing between the Higgs and radion modes. Finally, we calculate the production cross section of the radion at the LHC and use the current Higgs searches to place constraints on the parameter space.
\end{abstract}
\vspace*{\fill}

\eject
\pagestyle{empty}
\setcounter{page}{1}
\setcounter{footnote}{0}
\pagestyle{plain}

\section{Introduction}
For more than a decade extra dimensions have provided new insights into the hierarchy problem of 
the Standard Model (SM). By allowing fields to propagate in the extra dimensions the fundamental
cutoff scales can be lowered. This was first considered in the Arkani-Hamed, Dimopoulos, Dvali (ADD) model~\cite{ArkaniHamed:1998rs}, 
where if SM fields are confined to the boundary and only gravity propagates in the extra dimension, 
the fundamental scale of quantum gravity can be near the TeV scale. Assuming that the extra dimensions 
are stabilised, the weakness of gravity can then be attributed to the large volume occupied by the extra dimensions. 
An alternative possibility is to consider a warped extra dimension as in the Randall-Sundrum 
(RS) model~\cite{Randall:1999ee}. A warp factor then controls the scales of the four-dimensional (4D) Minkowski slices, 
causing the local cutoff to depend on the location in the extra dimension. 
In this case the hierarchy problem is addressed by choosing the Higgs to have a TeV cutoff scale.
This solution crucially relies on stabilising the extra dimension between the branes without
any tuning of parameters. The Goldberger-Wise mechanism~\cite{Goldberger:1999uk}, 
which introduces a bulk scalar field with boundary potentials, provides such a way to
stabilise the interbrane distance.

Recently a new solution generated by stacks of NS-5 branes in string theory was studied to also 
address the hierarchy problem. In the decoupling limit the  NS-5 branes give rise to Little String 
Theory -- a strongly-coupled nonlocal theory in six dimensions with no (apparent) Lagrangian 
description~\cite{Berkooz:1997cq,Seiberg:1997zk}. However by holography~\cite{Maldacena:1997re} 
the seven-dimensional gravity dual description is a linear dilaton background which can be used to study
the phenomenology of TeV Little String Theory~\cite{Antoniadis:2001sw}. An effective five-dimensional 
(5D) description can be obtained by compactifying two extra dimensions. This leads to the 5D linear dilaton 
model~\cite{Antoniadis:2011qw} which is compactified on a $\mathbb{Z}_2$ orbifold, with a 3-brane
placed at each of the endpoints. The 5D linear dilaton model contains features of both the ADD and RS models, 
including a large volume bulk like in the ADD model, and a warp factor like in the RS model. Moreover, 
phenomenologically the graviton Kaluza-Klein (KK) spectrum is distinct, consisting of a $\sim$ TeV mass 
gap followed by a near continuum of KK resonances separated by only $\sim$ 30 GeV~\cite{Antoniadis:2011qw,Baryakhtar:2012wj}. 

However, just like in previous solutions to the hierarchy problem involving extra dimensions, the issue
of stabilising the extra dimension is important. In this paper we investigate the stabilisation of the 
5D linear dilaton model and study the resulting radion phenomenology. An interesting feature of this 
model is that it already comes equipped with a scalar field, the dilaton, that can be used to stabilise 
the extra dimension. By adding boundary potentials, as in the Goldberger-Wise mechanism, a 
stabilising potential can then be obtained. The usual simplifying assumption, following from
Ref.~\cite{Goldberger:1999uk}, is to consider infinite boundary mass terms. However in our analysis
we will relax this condition and wherever possible consider finite, or else finite but large boundary mass terms, similar to
the general analysis considered in Ref~\cite{Kofman:2004tk, Gherghetta:2011rr}. Apart from the
usual radion couplings to the trace of the energy-momentum tensor, this leads to new radion 
couplings to SM fields that are confined to the brane. We present the Feynman rules and use 
them to study the radion phenomenology in this setup.  

Similar to the graviton KK spectrum, the radion mass spectrum consists of a near continuum 
of resonances spaced by $\sim$ 30 GeV together with a massive radion zero mode. The radion 
zero-mode and lowest KK mode masses can be parametrically lighter than the $\sim$ TeV mass gap, by suitably choosing 
the boundary mass parameters. The radion couples to the trace of the energy-momentum tensor
in the usual way, but in addition there is a new direct coupling between the dilaton and the SM fields.
The typical strength of these couplings is of order (10 TeV)$^{-1}$, with the dilaton coupling
being further suppressed by a factor inversely proportional to the boundary mass term. The SM fields have larger couplings to the radion 
compared to the near-continuum, and therefore the zero mode will be the first observable mode.

Furthermore, as is well known, the radion can kinetically mix with the Higgs boson via a Higgs-curvature 
interaction, but since the typical coupling strength is of order (10 TeV)$^{-1}$ there are negligible effects
on Higgs phenomenology. However, the production and decay of the radion leads to observable signals
at the LHC. In particular due to the additional direct coupling between the radion and gauge boson 
kinetic terms, the branching fraction to $\gamma\gamma$ can be significantly enhanced. Since the decays
are similar to that of the Higgs boson, current Higgs searches are then used to constrain the parameter 
space of the model. 

The outline of our paper is as follows. In Section 2 we briefly review the 5D linear dilaton
model of Ref.~\cite{Antoniadis:2011qw}. In Section 3 we identify the radion and solve 
for the KK mass spectrum in the limit of large but finite boundary mass terms. The radion 
couplings to SM fields are computed in Section 4 where the effects of mixing from a 
Higgs-curvature interaction are also included. The phenomenology of the radion is
presented in Section 5 where the radion decay widths, branching fractions and production
at the LHC are studied. In particular recent LHC results are used to constrain the parameter space
of the model. Finally in Section 6 we present our conclusion. Further details of the computations
are included in the Appendices as well as the Feynman rules for the radion couplings.

\section{Linear Dilaton Model}
The set-up of the linear dilaton model is similar to the RS scenario, with a finite extra dimension, 
$z$, compactified on a $\mathbb{Z}_2$ orbifold, except that we will take the TeV scale to be the 
fundamental scale and the 4D Planck mass as the derived scale. There are two 3-branes located at $z=0$ and $z=r_c$, with the SM confined to the $z=0$ brane. We also introduce a 5D bulk scalar field known as the dilaton. In the 
Einstein frame we have the following action~\cite{Antoniadis:2011qw}
\begin{equation}
\label{eq:action}
S_{bulk}={\int}d^5x\sqrt{-g}\left[M^3\left(\mathcal{R}-\frac{1}{3}({\nabla}\varphi)^2\right)-e^{\frac{2}{3}\varphi}\Lambda\right],
\end{equation}
\begin{equation}
\label{eq:bdyaction}
S_{vis(hid)}={\int}d^4x\sqrt{-\hat{g}}e^{\frac{1}{3}\varphi}\left(\mathcal{L}_{vis(hid)}-V_{vis(hid)}\right),
\end{equation}
where $\varphi$ is the dilaton field and $M$ is the fundamental scale, which is of order the TeV scale. In the conformal coordinate, $z$, the background solutions to the field equations are given by 
\begin{equation}
\phi(z)=\alpha{\vert}z\vert,
\end{equation}
\begin{equation}\label{eq:metric}
ds^2=e^{-\frac{2}{3}{\alpha}{\vert}z\vert}\left(\eta_{\mu\nu}dx^{\mu}dx^{\nu}+dz^2\right),
\end{equation}
where $\phi(z)$ denotes the background value of the dilaton field, $\varphi$, and $\vert\alpha\vert<\frac{3M}{2\sqrt{7}}$ is required to ensure the 5D curvature is smaller than the fundamental scale. There are also the following constraints
\begin{equation}
\label{eq:LDbc}
\Lambda=-M^3\alpha^2,{\quad}V_{vis}=-V_{hid}=4{\alpha}M^3,
\end{equation}
which correspond to the usual tuning of the 4D cosmological constant.

Taking the curvature term in the action \eqref{eq:action} and integrating over the fifth dimension yields the effective 4D Planck mass, 
\begin{equation}
M_{Pl}^2=2\int^{r_c}_{0}dz\,e^{-{\alpha}{\vert}z\vert}M^3=-2\frac{M^3}{\alpha}\left(e^{-\alpha r_c}-1\right).
\end{equation}
We then clearly require that $\alpha<0$ and $\vert{\alpha}r_c\vert\sim70$ in order to obtain the required value for $M_{Pl}$.

In order to allow a straightforward comparison with the RS model, we can perform the following coordinate transformation $dy=e^{-\frac{1}{3}{\alpha}z}dz$. In the new coordinate, the metric becomes
\begin{equation} \label{eq:metric-proper}
ds^2=\left(1+\frac{\vert{\alpha}y\vert}{3}\right)^2\eta_{\mu\nu}dx^{\mu}dx^{\nu}+dy^2.
\end{equation} 
We can now see that this model exhibits a power-law warping as opposed to the exponential warping of the RS model. The size of the extra dimension in this case must therefore be substantially larger than in the RS model in order to obtain the necessary hierarchy. Taking $\vert\alpha\vert\sim\mathrm{TeV}$ and enforcing the correct value for $M_{Pl}$ allows us to determine the proper length of the extra dimension, giving $y_c\sim10\ \mathrm{nm}$. This compares with a proper length
of $\sim 0.1$ mm in the ADD model (for two extra dimensions) and $\sim 10^{-31}$ cm in the RS1 model.

The linear dilaton model has an interesting spectrum of graviton KK modes. There is a single massless mode, which has a flat profile in the extra dimension and is identified with the usual 4D graviton.
In addition, there is a KK tower of excited states, with a mass spectrum given by
\begin{equation} \label{eq:graviton-spectrum}
m_n^2=\frac{\alpha^2}{4}+\left(\frac{n\pi}{r_c}\right)^2,
\end{equation} 
where $n=1,2,3,...\ $. Of particular interest is a large mass gap between the massless graviton and the first of the KK modes, which are then closely spaced. The mass gap is determined by the curvature scale, $\alpha$, and for a TeV mass gap we find that $r_c\sim(30\ \mathrm{GeV})^{-1}$ and the KK modes essentially form a continuum of states. The graviton KK modes are localised near the SM ($z=0$) brane, and their couplings to the Standard Model fields are given in Ref. \cite{Antoniadis:2011qw}. They are found to couple to the Standard Model fields with a strength of order $(\mathrm{80\ TeV})^{-1}$ for the lowest mode.

\subsection{Stabilisation}
In order to truly solve the hierarchy problem the interbrane distance must be stabilised. In the linear dilaton solution of \cite{Antoniadis:2011qw} this can be done by using a variation of the Goldberger-Wise mechanism. Unlike the Randall-Sundrum scenario, the linear dilaton  model naturally includes a bulk scalar field, the dilaton, that can in fact play the role of the stabilising field once it acquires a vacuum expectation value on the branes.

We begin by taking boundary potentials of the form
\begin{equation}\label{eq:b-potential}
V_{vis(hid)}(\varphi)={\pm}\lambda_{vis(hid)}+\mu_{vis(hid)}M^3(\varphi-\phi_{vis(hid)})^2,
\end{equation}
where $\lambda_{vis(hid)}$, $\mu_{vis(hid)}$ and $\phi_{vis(hid)}$ are constants. At this point it is useful to formulate the model in terms of the solution generating method of \cite{DeWolfe:1999cp}, by considering the superpotential $W(\varphi)=W_0e^{\frac{\varphi}{3}}$, where $W_0$ is a constant. The background field equations include two junction conditions at the boundaries, which can now be expressed as
\begin{gather}
W(\phi)={\pm}e^{\frac{\phi}{3}}V_{vis(hid)}(\phi),\\
\frac{{\partial}W(\phi)}{\partial\phi}=\pm\frac{\partial}{\partial\phi}\left(e^{\frac{\phi}{3}}V_{vis(hid)}(\phi)\right).
\end{gather}
Using \eqref{eq:b-potential}, these equations can only be simultaneously satisfied if $\phi=\phi_{vis(hid)}$ on the boundaries. 
Thus we see that $\varphi$ has developed a non-zero vacuum expectation value on the branes as a result of imposing the boundary potentials \eqref{eq:b-potential} (setting $\mu_{vis(hid)}=0$ also satisfies the conditions, however this simply reproduces the unstabilised case). It is worth noting that an additional linear term can be added to the boundary potential, however this simply shifts the value of the VEV and the potential can always be rewritten in the form \eqref{eq:b-potential}. The requirement that $\phi=\phi_{vis(hid)}$ on the boundaries then gives 
\begin{equation}
\label{eq:bc1}
\lambda_{vis(hid)}=W_0=4\sqrt{-\Lambda M^3},
\end{equation}
where we have used the definition of the bulk potential from the action \eqref{eq:action}. The condition \eqref{eq:bc1} corresponds to the usual tuning of the 4D cosmological constant in \eqref{eq:LDbc} necessary to obtain a flat brane solution. 

The background solution for the bulk scalar is once again given by
\begin{equation}
\phi(z)=\alpha{\vert}z\vert+\bar\phi,
\end{equation}
where here we have included the integration constant, $\bar\phi$, which was previously taken to be zero. Imposing the boundary conditions $\phi=\phi_{vis}$ and $\phi=\phi_{hid}$ on the $z=0$ and $z=r_c$ branes respectively, we then find that
\begin{gather}
\bar\phi=\phi_{vis},\\
{\alpha}r_c=\phi_{hid}-\phi_{vis}.
\end{gather}
The second expression clearly shows that we have indeed stabilised the brane separation, $r_c$. In order to achieve the correct hierarchy, we previously required that $\vert{\alpha}r_c\vert\sim70$. This can be achieved without the need for any extreme fine-tuning of the parameters. In the proper coordinates \eqref{eq:metric-proper} this corresponds to stabilising the interbrane distance at $y_c\sim10^{10}\vert\alpha\vert^{-1}$, much larger than in RS models. Finally, it is important to note that since $\phi_{vis}$ appears in the exponential dilaton factor in the boundary action \eqref{eq:bdyaction}, we require that $\phi_{vis}=0$ in order to successfully reproduce the Standard Model. This is the case we shall consider from now on.

\section{Identification of the Radion}\label{sec:radion}
In this section we identify the physical radion and KK modes and calculate the mass spectrum. The scalar fluctuations about the background solutions are denoted
\begin{gather}
ds^2=e^{-\frac{2}{3}{\alpha}{\vert}z\vert}\left[(1+2\Psi)\eta_{\mu\nu}dx^{\mu}dx^{\nu}+(1+2\Phi)dz^2\right],\label{eq:decomposition}\\
\varphi(x,z)=\phi(z)+\delta\phi(x,z).
\end{gather}
We then obtain the linearised Einstein equations for the perturbations, which include the following constraint equations
\begin{equation}\label{eq:constraint}
\begin{split}
&\Phi+\frac{3}{\alpha}\Psi'+\frac{1}{3}\delta\phi=0,\\
&\Phi=-2\Psi,
\end{split}
\end{equation}
where primes denote differentiation with respect to $z$. The presence of these non-dynamical equations reflects the fact that the gravitational and dilaton scalar perturbations are coupled and the action must be diagonalised in order to identify the physical degree of freedom of the system. Using these constraint equations the dynamical equation can be expressed as
\begin{equation}\label{eq:eom}
\left[\Box+\frac{d^2}{dz^2}-\frac{\alpha^2}{4}\right]\left(e^{-\frac{1}{2}{\alpha}z}\Phi\right)=0.
\end{equation}
There are also two junction conditions at the boundaries. The first is obtained by simply evaluating \eqref{eq:constraint} at the boundaries, while the second contains additional physical information and is given by
\begin{equation}
\delta\phi=\pm\frac{4M^3e^{\frac{1}{3}{\alpha}z}}{\frac{\partial^2}{\partial\phi^2}\left(3e^{\frac{\phi}{3}}V_{vis(hid)}(\phi)\right)}\left(\delta\phi'-{\alpha}\Phi\right),
\end{equation}
One of the nice features of this model is that the equations of motion for the perturbations can be solved analytically. While the boundary conditions are non-trivial, they become considerably simpler in the limit $\frac{\partial^2}{\partial\phi^2}(e^{\frac{\phi}{3}}V)\rightarrow\infty$, which forces $\delta\phi\rightarrow0$ on the branes. This is the limit most often considered in the literature, including the original Goldberger-Wise case. However, this will be insufficient for our purposes since many of the interesting phenomenological features of this model arise due to the non-zero coupling of the dilaton field to the SM Lagrangian on the brane, as shall be seen in sections~\ref{sec:coupling} and \ref{sec:decays}.

We begin by performing a KK decomposition
\begin{equation}
\Phi(x^\mu,z)=\sum_{n=0}^\infty\Phi_n(z)Q_n(x^\mu),
\end{equation}
where $Q_n(x^\mu)$ are the 4D Kaluza-Klein modes and satisfy the Klein-Gordon equation, and $\Phi_n$ are the profiles in the fifth dimension. The equation of motion is then simply given by
\begin{equation} \label{eq:evalue_eq}
\left[\frac{d^2}{dz^2}+m_n^2-\frac{\alpha^2}{4}\right]\left(e^{-\frac{1}{2}{\alpha}z}\Phi_n\right)=0,
\end{equation}
with the boundary condition
\begin{equation}
\frac{3}{2\alpha}\Phi_n'-\Phi_n=\frac{2}{2\alpha\pm9\mu_{vis(hid)}}\left(\frac{9}{2\alpha}\Phi_n''-3\Phi_n'-\alpha\Phi_n\right).
\end{equation}
We obtain the following solution for $\Phi_n$,
\begin{equation} \label{eq:wavefunction}
\Phi_n(z)=N_ne^{\frac{1}{2}{\alpha}z}\left[\sin(\beta_nz)-\frac{6\beta_n\mu_{vis}}{4\beta_n^2+\alpha(\alpha-\mu_{vis})}\cos(\beta_nz)\right],
\end{equation}
where we have defined $\beta_n^2\ {\equiv}\ m_n^2-\frac{\alpha^2}{4}$, $N_n$ is an overall normalisation factor, and we have used the boundary condition at $z=0$ to fix one of the constants. The mass spectrum is then obtained by evaluating the boundary condition at the $z=r_c$ boundary. The complete mass spectrum can be determined analytically by taking $\mu_{vis(hid)}$ to be large and solving as an expansion in $\vert\alpha\vert/\mu_{vis(hid)}$. However, an exact expression can also be obtained for the radion and lowest KK mode in the limit $\vert{\alpha}r_c\vert\gg1$. We then find that 
\begin{equation} \label{eq:rad-mass}
m_{rad}^2=\frac{\alpha^2}{4}-\frac{\alpha^2}{16\epsilon_{vis}^2}\left(3-\sqrt{9+4\epsilon_{vis}+4\epsilon_{vis}^2}\right)^2,\qquad0<\epsilon_{vis}<\infty,
\end{equation}
where $\epsilon_{vis(hid)}\equiv\vert\alpha\vert/\mu_{vis(hid)}$ and $\mu_{vis(hid)}>0$ is required to ensure there are no tachyonic modes. Note that dependence on $\epsilon_{hid}$ in \eqref{eq:rad-mass} is suppressed in the $\vert{\alpha}r_c\vert\gg1$ limit. We also solve the boundary condition as an expansion in $\epsilon_{vis(hid)}$ to obtain the complete KK spectrum at first order
\begin{align}
m_{rad}^2 &=\frac{2}{9}\alpha^2\left(1-\frac{2\epsilon_{vis}}{9}\right),\\
m_n^2 &=\frac{\alpha^2}{4}+\frac{n^2\pi^2}{r_c^2}\left[1-\frac{4(4n^2\pi^2+3\vert{\alpha}r_c\vert^2)}{12n^2\pi^2\vert{\alpha}r_c\vert+\vert{\alpha}r_c\vert^3}\left(\epsilon_{vis}+\epsilon_{hid}\right)\right],
\end{align}
where $n=1,2,3,...\ $. Note that in the limit $\epsilon_{vis(hid)}\rightarrow0$, corresponding to $\mu_{vis(hid)}\rightarrow\infty$, we obtain the result given in \cite{Kofman:2004tk}. 

We see that this model also gives rise to a rather interesting mass spectrum for the scalar fluctuations. Both the radion and lowest KK mode have masses of order the curvature scale, but can be parametrically lighter depending on the values of $\epsilon_{vis(hid)}$. The spacing of the KK modes is then determined by the size of the extra dimension as we would expect. Once again we have $r_c\sim(30\ \mathrm{GeV})^{-1}$ and the KK spectrum forms a near continuum of modes above a mass of approximately half the curvature scale. 

\begin{figure}[H]
\begin{center}
\includegraphics[height=7cm]{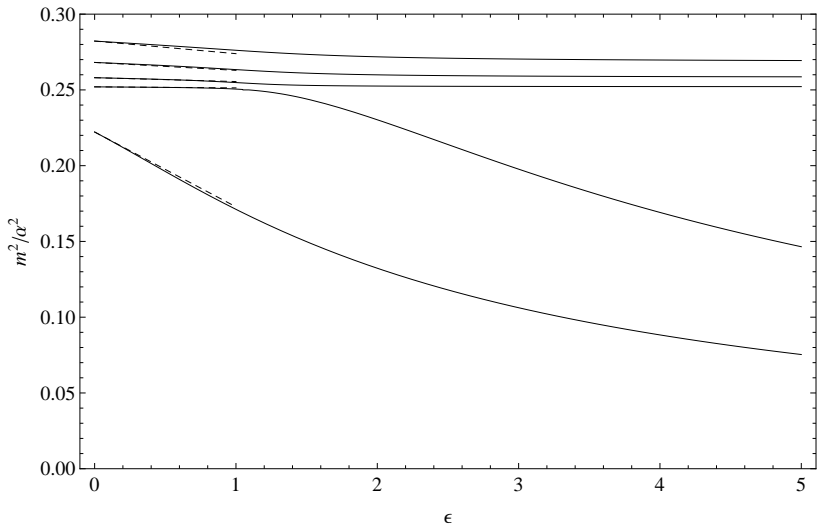}
\caption{The radion and KK mass spectrum as a function of $\epsilon$, where we have taken $\epsilon_{vis}=\epsilon_{hid}$. The lowest five modes are shown and the dashed lines denote the approximate solutions.}
\label{fig:mass-eps}
\end{center}
\end{figure}

The mass spectrum is plotted as a function of $\epsilon$ in Figure~\ref{fig:mass-eps}, where the KK modes have been determined numerically and we have taken $\epsilon_{vis}=\epsilon_{hid}$. Note however that as seen in \eqref{eq:rad-mass} the lowest mode is dependent only on $\epsilon_{vis}$, while the second lowest mode depends only on $\epsilon_{hid}$ for $\epsilon_{hid}\gtrsim1$. We find that the exact and approximate solutions are well matched when $\epsilon$ is small, justifying our expansion. Interestingly, we observe that both the radion and lowest KK mode are highly sensitive to the boundary mass terms and as $\epsilon$ is increased there are two distinct modes below the near continuum of KK states. In the $\epsilon\rightarrow\infty$ limit, corresponding to the unstabilised case with $\mu_{vis(hid)}=0$, the KK spectrum is identical to the graviton spectrum \eqref{eq:graviton-spectrum} with a $\sim\ $TeV mass gap. We note that while the decomposition \eqref{eq:decomposition} becomes ambiguous when considering massless scalar modes, we expect two massless modes in this limit, where the additional massless mode arises due to the extra bulk scalar field (the dilaton).

\section{Coupling to the Standard Model}
\label{sec:coupling}
In this section we calculate the couplings of the radion and KK modes to the Standard Model fields. The relevant term in the boundary action is
\begin{equation}
S={\int}d^4x\sqrt{-\hat{g}}\, e^{\frac{\delta\phi}{3}}\mathcal{L}_{SM},
\end{equation}
where $\mathcal{L}_{SM}$ is the SM Lagrangian and $\hat{g}$ is the induced metric on the $z=0$ brane. We clearly see that the radion couples to the SM fields through the induced metric on the brane and also via the exponential dilaton factor. This dilaton factor is not present in the usual RS models and gives rise to an additional coupling between the radion and SM fields. Expanding the above action to first order in the perturbations, we obtain the following interaction action between the radion KK modes, $Q_n$, and the SM fields
\begin{equation}\label{eq:int_action}
S_{int}=\frac{1}{2}\sum_{n}\Phi_{n}(0){\int}d^4x\sqrt{-\hat{g}_0}\,Q_{n}\,\hat{g}_0^{\mu\nu}T_{\mu\nu}+\frac{1}{3} \sum_{n}\delta\phi_n(0) {\int}d^4x\sqrt{-\hat{g}_0}\,Q_{n}\,\mathcal{L}_{SM}\big\vert_{\hat{g}=\hat{g}_0}
\end{equation}
where $\hat{g}_0$ is the induced background metric on the brane and $T_{\mu\nu}$ is the Standard Model energy-momentum tensor. The first term in \eqref{eq:int_action} is the usual coupling of the radion to the trace of the energy momentum tensor, while the second term arises due to the presence of the dilaton and produces a direct coupling between the radion and SM Lagrangian. This additional coupling can have a significant effect on the radion phenomenology, as we shall see in section~\ref{sec:decays}. 

For convenience we define the following coupling constants
\begin{equation}
\frac{\kappa_{\Phi,n}}{M}\equiv\frac{\Phi_n(0)}{2}, \qquad\frac{\kappa_{\phi,n}}{M}\equiv\frac{\delta\phi_n(0)}{3}.
\end{equation}
The couplings can now be determined by correctly normalising the solutions for the perturbations obtained in section~\ref{sec:radion}. As we found in section~\ref{sec:radion}, the couplings for the radion can be determined analytically in the $\vert{\alpha}r_c\vert\gg1$ limit, while the KK couplings can only be obtained as an expansion in $\epsilon_{vis(hid)}$. The exact expression for the radion coupling is non-trivial and does not provide additional insight, so here we give only the approximate solutions. The couplings to the energy-momentum tensor are given by
\begin{align} 
\vert\kappa_{\Phi,rad}\vert&=\frac{1}{6\sqrt{2}}\sqrt{\frac{\vert\alpha\vert}{M}}\left(1+\frac{4}{9}\epsilon_{vis}\right),\\
\vert\kappa_{\Phi,n}\vert&=\frac{4n\pi}{\sqrt{6}\vert{\alpha}r_c\vert^{3/2}}\sqrt{\frac{\vert\alpha\vert}{M}}\left(1-\epsilon_{vis}\right),\label{eq:couplings}
\end{align}
where we have also taken $\vert{\alpha}r_c\vert\gg1$, since the expressions simplify significantly in this limit. We see that the coupling strength is largely determined by the fundamental scale, $M$ and also the ratio $\alpha/M$. Additionally, the couplings for the KK modes are suppressed by a factor of $\vert{\alpha}r_c\vert^{3/2}$ relative to the radion mode. Hence, we always expect the single radion mode to be observed first by experiments, even when the radion mass is close to that of the KK near continuum or when there are two light modes. It is also interesting to note that the couplings for the KK modes increase with $n$, however this remains true only for small $n$ since \eqref{eq:couplings} is no longer valid when $n^2\sim\vert{\alpha}r_c\vert$.

Using the constraint equation for the perturbations \eqref{eq:constraint} evaluated at the $z=0$ brane we can also determine the couplings to the SM Lagrangian, which are given by
\begin{align}
\vert\kappa_{\phi,rad}\vert&=\frac{\sqrt{2}}{27}\sqrt{\frac{\vert\alpha\vert}{M}}\epsilon_{vis},\\
\vert\kappa_{\phi,n}\vert&=\frac{2n\pi}{\sqrt{6}\vert{\alpha}r_c\vert^{3/2}}\sqrt{\frac{\vert\alpha\vert}{M}}\epsilon_{vis}.
\end{align}
We see that unlike $\kappa_\Phi$ above, these couplings do not include a zeroth order term, consistent with the fact that $\delta\phi\rightarrow0$ in the $\mu_{vis}\rightarrow\infty$ limit. The couplings to the SM Lagrangian are therefore suppressed by an additional factor of $\vert\alpha\vert/\mu_{vis}$, which must be small in order to justify our expansion. Once again, the KK modes are further suppressed relative to the radion mode.

\begin{figure}[H]
\begin{center}
\includegraphics[height=7cm]{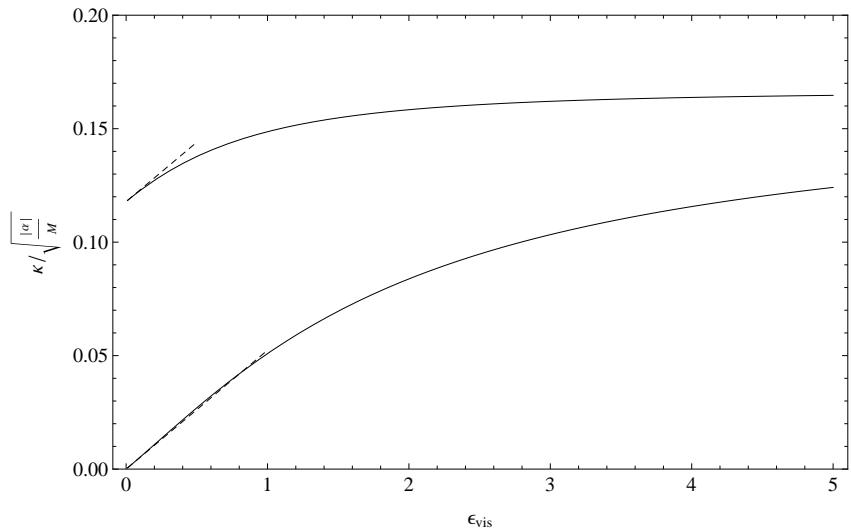}
\caption{The radion couplings as a function of $\epsilon_{vis}$. The upper and lower curves correspond to $\kappa_\Phi$ and $\kappa_\phi$ respectively. The dashed lines denote the approximate solutions.}
\label{fig:couplings}
\end{center}
\end{figure}

Figure~\ref{fig:couplings} shows the radion couplings as a function of $\epsilon_{vis}$, where the upper and lower curves correspond to $\kappa_\Phi$ and $\kappa_\phi$ respectively. Once again we see that the approximate and exact solutions are in good agreement when $\epsilon_{vis}$ is small. We also find that $\kappa_\Phi>\kappa_\phi$, even for larger values of $\epsilon_{vis}$ where our approximate solution is no longer valid. Taking $M\sim\mathrm{TeV}$, $\vert\alpha\vert/M\sim0.5$ and $\epsilon_{vis}=3$, we find that $\kappa_{\Phi,rad}\sim0.11$ and $\kappa_{\Phi,rad}/M$ will be of order $\sim(\mathrm{9\ TeV})^{-1}$. Similarly, $\kappa_{\phi,rad}\sim0.07$ and $\kappa_{\phi,rad}/M$ will be of order $\sim(\mathrm{14\ TeV})^{-1}$. Note also that while the overall sign of the couplings is undetermined by the normalisation condition, the relative sign is fixed giving $\kappa_\Phi/\kappa_\phi>0$.

\subsection{Higgs-curvature interaction}\label{sec:higgs-curvature}
An interesting situation arises for scalar fields propagating on the brane, since they can couple non-minimally to gravity. This allows us to introduce the following additional Higgs-curvature interaction term to the 4D effective action.
\begin{equation}\label{eq:higgs-curvature}
S_\xi={\int}d^4x\sqrt{-\hat{g}}e^{\frac{\delta\phi}{3}}\xi\mathcal{R}(\hat{g})H^{\dagger}H,
\end{equation}
where $\mathcal{R}(\hat{g})$ is the Ricci scalar obtained from the 4D induced metric on the brane. For an arbitrary scalar field $\psi$, we could also in principle include the linear coupling $\mathcal{R}(\hat{g})\psi$, however in the case of the Higgs this is forbidden by the gauge symmetry. As we shall see, this additional interaction term results in a mixing between the Higgs and the radion. The analysis given below follows that performed for the RS case in \cite{Giudice:2000av, Csaki:2000zn}, however here we include the additional effects of the dilaton field.

The 4D Ricci scalar is calculated from the induced metric to be
\begin{equation}
\mathcal{R}(\hat{g})=-6(1-\Phi)^{-3/2}\eta^{\mu\nu}\partial_\mu\partial_\nu(1-\Phi)^{1/2},
\end{equation}
which results in the following Lagrangian for the interaction
\begin{equation}
\mathcal{L}_\xi=-6{\xi}e^{\frac{\delta\phi}{3}}(1-\Phi)^{1/2}\eta^{\mu\nu}\partial_\mu\partial_\nu(1-\Phi)^{1/2}H^{\dagger}H.
\end{equation}
Now, after expanding the Higgs field around its vacuum expectation value, $H=(h+v)/\sqrt{2}$, where $v\simeq246\ $GeV, to quadratic order in the fields we have
\begin{equation}
\mathcal{L}_\xi=-3\xi\left(\frac{1}{4}v^2\Phi\Box\Phi-\frac{1}{6}v^2\delta\phi\Box\Phi-vh\Box\Phi\right).
\end{equation}
Finally, inserting the KK expansion for the fields and considering only the radion mode, $r$, we obtain the following complete Higgs-radion Lagrangian
\begin{equation}
\mathcal{L}=-\frac{1}{2}h{\Box}h-\frac{1}{2}m_h^2h^2-\frac{1}{2}\left[1+\frac{6\xi\kappa_{\Phi}v^2}{M^2}(\kappa_\Phi-\kappa_\phi)\right]r{\Box}r-\frac{1}{2}m_r^2r^2+\frac{6{\xi}\kappa_{\Phi}v}{M}h{\Box}r,
\end{equation}
where $m_h$, $m_r$ are the Higgs and radion masses respectively, in the $\xi=0$ limit. The $\xi$ term then clearly introduces a kinetic mixing between the Higgs and radion fields. The Lagrangian must therefore be diagonalised to identify the physical mass eigenstates of the system. We begin by diagonalising the kinetic terms via the following transformation
\begin{equation}
h=h'+\frac{6{\xi}\kappa_{\Phi}v}{{\Omega}M}r',\qquad r=\frac{r'}{\Omega},
\end{equation}
where
\begin{equation}
\Omega^2=1+\frac{6\xi\kappa_{\Phi}v^2}{M^2}\bigg((1-6\xi)\kappa_\Phi-\kappa_\phi\bigg).
\end{equation}
This in fact also enables us to place a constraint on the parameter $\xi$, since $\Omega^2$ must be positive in order to ensure that the radion mass term retains the correct sign and we avoid encountering a ghost radion field. We therefore require that $\xi$ lies in the range
\begin{equation}
\frac{1}{12\kappa_\Phi}\left(\rho-\sqrt{\rho^2+\frac{4M^2}{v^2}}\right)\leq\xi\leq\frac{1}{12\kappa_\Phi}\left(\rho+\sqrt{\rho^2+\frac{4M^2}{v^2}}\right),
\end{equation}
where $\rho\equiv\kappa_\Phi-\kappa_\phi$.
Taking $\kappa_\Phi=0.11$, $\kappa_\phi=0.07$ and $M=1\ $TeV, we find that this places a limit of $-6.1<\xi<6.2$.

So far, we have successfully diagonalised the kinetic terms in the Lagrangian, however this now introduces a mixing in the mass matrix, which must be diagonalised via the rotation
\begin{equation}
\begin{split}
h' &=\cos{\theta}h_m+\sin{\theta}r_m,\\
r' &=-\sin{\theta}h_m+\cos{\theta}r_m,
\end{split}
\end{equation}
where $h_m$, $r_m$ are the physical mass eigenstates and the mixing angle is given by
\begin{equation}
\tan{2\theta}=\frac{12{\xi}\kappa_{\Phi}v{\Omega}}{M}\frac{m_h^2}{m_r^2-m_h^2\left(\Omega^2-\frac{36\xi^2\kappa_{\Phi}^2v^2}{M^2}\right)}.
\end{equation}
We can now express the gauge eigenstates in terms of the mass eigenstates as
\begin{equation}\label{eq:mixing_full}
\begin{split}
h &=\left(\cos\theta-\frac{6{\xi}\kappa_{\Phi}v}{{\Omega}M}\sin\theta\right)h_m+\left(\sin\theta+\frac{6{\xi}\kappa_{\Phi}v}{{\Omega}M}\cos\theta\right)r_m,\\
r &=-\frac{\sin\theta}{\Omega}h_m+\frac{\cos\theta}{\Omega}r_m.
\end{split}
\end{equation}
We can also calculate the corresponding mass eigenvalues, which are given by
\begin{eqnarray}
m_{r_m,h_m}^2&=&\frac{1}{2\Omega^2}\left[m_r^2+m_h^2\left(1+\frac{6{\xi}\kappa_{\Phi}^2v^2}{M^2}\right)
\right.\nonumber\\
&&\qquad\left.
\pm\left(\left(m_r^2-m_h^2\left(1+\frac{6{\xi}\kappa_{\Phi}^2v^2}{M^2}\right)\right)^2+\frac{144\xi^2\kappa_{\Phi}^2v^2m_r^2m_h^2}{M^2}\right)^{1/2}\right]
\end{eqnarray}

In the following section we shall only be considering the radion interactions described in \eqref{eq:int_action} and will not be concerned with higher-dimensional operators, which are suppressed by higher powers of $M$. It therefore makes sense to expand the above results to leading order in $1/M$. The mixing angle then becomes
\begin{equation}
\label{eq:mix-angle}
\tan{2\theta}\approx\frac{12{\xi}\kappa_{\Phi}v}{M}\frac{m_h^2}{m_r^2-m_h^2}.
\end{equation}
Now, provided that $m_r\displaystyle{\not}{\approx}m_h$, the mixing angle is small and our results simplify to 
\begin{equation}\label{eq:mixing}
\begin{split}
h &=h_m+\frac{6{\xi}\kappa_{\Phi}v}{M}\frac{m_r^2}{m_r^2-m_h^2}r_m,\\
r &=-\frac{6{\xi}\kappa_{\Phi}v}{M}\frac{m_h^2}{m_r^2-m_h^2}h_m+r_m.
\end{split}
\end{equation}

This mixing can in principle have an interesting effect on the Higgs phenomenology, since a strong mixing would lead to a reduced cross section for the Higgs. However, \eqref{eq:mixing} shows that the extent of the mixing is largely determined by the factor $\kappa_{\Phi}v/M$ and is therefore small, except in the case when $m_r$ is very near to $m_h$, or if we take $\xi$ to be large. While the mixing may not have any significant effect on the Higgs, it still plays an important role in determining the radion phenomenology, since the extent of the mixing is comparable to the size of the radion couplings.

When $m_r{\approx}m_h$, \eqref{eq:mixing} is no longer valid and we must include higher order terms in the expression for the mixing angle \eqref{eq:mix-angle} or else consider the full expressions in \eqref{eq:mixing_full}. In this case we find that the mixing becomes large and the two mass eigenstates are now essentially part Higgs and part radion. The result is two states with closely spaced masses, both of which interact in a similar way to the SM Higgs but with a reduced cross section.

We also note that at leading order our result does not depend on $\kappa_\phi$ and the dilaton does not contribute significantly to the Higgs-radion mixing. Our results are then equivalent to those for the RS case given in \cite{Giudice:2000av, Csaki:2000zn}. In this limit, the mass eigenvalues also simply reduce to $m_r$, $m_h$. The above analysis can also be performed with the inclusion of the radion KK modes
and is given in Appendix~\ref{sec:KKmixing}. The mixing effect is sub-dominant compared to the radion only case.

\subsection{SM-Radion interactions}
As we have shown, the radion couples to the trace of the Standard Model energy-momentum tensor as well as directly to the SM Lagrangian. We now derive in detail the radion interactions with the Standard Model fields
and present the corresponding Feynman rules for the interaction terms in Appendix~\ref{sec:feynrules}. 

Following on from our discussion of the previous section, we consider the case where there is a mixing between the radion and Higgs fields and express the interactions in terms of the physical mass eigenstates. In order to simplify the expressions, we shall use the following general formulae relating the gauge and mass eigenstates, where the values of the coefficients can be read off from \eqref{eq:mixing_full}:
\begin{equation}
\begin{split}
h&=a_0h_m+a_1r_m,\\
r&=b_0h_m+b_1r_m.
\end{split}
\end{equation}
We shall also restrict ourselves to interactions up to quadratic order in the SM fields, since these will be the dominant processes when considering the radion decay modes in the following section. The Standard Model energy-momentum tensor is given by
\begin{multline}\label{eq:T_mm}
T_\mu^\mu=-2m_W^2W_\mu^+W^{-\mu}-m_Z^2Z_{\mu}Z^\mu+2m_h^2h^2-\partial_{\mu}h\partial^{\mu}h\\
+\sum_f\frac{3}{2}\partial_\mu(\bar{\Psi}i\gamma^\mu\Psi)-3\bar{\Psi}i\gamma^\mu\partial_\mu\Psi+4m_\Psi\bar{\Psi}\Psi.
\end{multline}

We begin by considering the interactions of the radion with the massive gauge bosons, which are given in the unitary gauge by
\begin{equation}
\mathcal{L}_{int}=-\frac{b_1\kappa_\phi}{4M}r_m\left(2W^\dagger_{\mu\nu}W^{\mu\nu}+Z_{\mu\nu}Z^{\mu\nu}\right)+\left(\frac{b_1}{M}(\frac{\kappa_\phi}{2}-\kappa_\Phi)+\frac{a_1}{v}\right)r_m\left(2m_W^2W^\dagger_{\mu}W^\mu+m_Z^2Z_{\mu}Z^\mu\right).
\end{equation}
We see that as a result of the dilaton coupling to the SM Lagrangian, there is now a coupling between the radion and the gauge boson kinetic terms, which will become important at large momenta. As we shall see, this additional coupling is particularly significant for the massless gauge fields.

We now consider the interactions with the SM fermions which are given by
\begin{equation}
\mathcal{L}_{int}=\frac{b_1}{M}\left(\kappa_\phi-3\kappa_\Phi\right)r_m\bar{\Psi}i\gamma^\mu\partial_\mu\Psi-\frac{3b_1\kappa_\Phi}{2M}\partial_{\mu}r_m\bar{\Psi}i\gamma^\mu\Psi+m_\Psi\left(\frac{b_1}{M}(4\kappa_\Phi-\kappa_\phi)-\frac{a_1}{v}\right)r_m\bar{\Psi}{\Psi}.
\end{equation}
However, in most of the cases we shall be considering it will be sufficient to restrict ourselves to on-shell fermions. Then using the Dirac equation we can significantly simplify the above interaction and obtain
\begin{equation}
\mathcal{L}_{int}=m_\Psi\left(\frac{b_1\kappa_\Phi}{M}-\frac{a_1}{v}\right)r_m\bar{\Psi}{\Psi}.
\end{equation}
Of course, the Dirac Lagrangian vanishes after using the equations of motion and our interaction term is therefore independent of $\kappa_\phi$.

There are additional considerations when looking at the radion interactions with the Higgs boson. As a result of the mixing between the radion and the Higgs, the interactions will now be dependent on the radion potential as well as the Higgs self-interaction terms. In addition to this, there are interaction terms which arise from the Higgs-curvature term \eqref{eq:higgs-curvature}. Following the analysis in section~\ref{sec:higgs-curvature}, but here considering terms which are cubic in the fields, gives the following interaction term to leading order in $1/M$
\begin{equation}
\mathcal{L}_{int}^\xi=\frac{6{\xi}\kappa_{\Phi}a_0^2b_1}{M}\left(\partial_{\mu}h_m\partial^{\mu}h_m+h_m{\Box}h_m\right)r_m.
\end{equation}
Now putting this together with the other interaction terms we obtain
\begin{multline}
\mathcal{L}_{int}=\frac{a_0^2b_1}{M}\bigg((6\xi-1)\kappa_\Phi+\frac{\kappa_\phi}{2}\bigg)r_m\partial_{\mu}h_m\partial^{\mu}h_m+\frac{6{\xi}\kappa_{\Phi}a_0^2b_1}{M}r_mh_m{\Box}h_m\\ +a_0^2m_h^2\left(\frac{b_1}{M}(2\kappa_\Phi-\frac{\kappa_\phi}{2})-\frac{3a_1}{2v}\right)r_mh_m^2,
\end{multline}
where the final term arises from the cubic Higgs self-interaction term in the SM Lagrangian. We have excluded the effects of the radion potential terms, which we shall assume to be small.

Finally we consider the interactions with the massless gauge bosons. Here we must also include 
the quantum effects of the scale anomaly, which is reviewed in \cite{Bae:2001id}.
The interaction with the gluons is given by
\begin{equation}
\mathcal{L}_{int}=-\frac{b_1}{M}\left(\frac{\kappa_\phi}{4}+\frac{\alpha_sb_{QCD}\kappa_\Phi}{8\pi}\right)r_m\mathcal{F}^a_{\mu\nu}\mathcal{F}_a^{\mu\nu},
\end{equation}
where $b_{QCD}=11-2n_f/3$ with $n_f=6$.
For the gluons it turns out that the direct coupling of the dilaton and the additional coupling due to the scale anomaly are of comparable strength due to the relatively large value of $\alpha_s$ and the suppression of $\kappa_\phi$ relative to $\kappa_\Phi$. 

The interaction with the photon is similarly
\begin{equation}
\mathcal{L}_{int}=-\frac{b_1}{M}\left(\frac{\kappa_\phi}{4}+\frac{\alpha_{EM}\kappa_\Phi}{8\pi}(b_2+b_Y)\right)r_mF_{\mu\nu}F^{\mu\nu},
\end{equation}
where $b_2=19/6$ and $b_Y=-41/6$. However in this case the anomalous contributions to $T_\mu^\mu$ are insignificant due to the much smaller value of $\alpha_{EM}$. We similarly ignored the effects of the scale anomaly when considering the $W$ and $Z$ gauge bosons above.

\section{Radion Decays and Production}\label{sec:decays}
\subsection{Decay widths}
From the interaction terms given above and the Feynman rules given in Appendix~\ref{sec:feynrules} we calculate the partial decay widths of the radion into $WW$, $ZZ$, $hh$, $\bar{f}f$, $\gamma\gamma$, and $gg$. We have taken our results to leading order in $1/M$ and used the expression \eqref{eq:mixing} for the Higgs-radion mixing. They are given by

\begin{align}
\Gamma(r{\rightarrow}W^+W^-)&=\frac{m_r^3}{16{\pi}M^2}\sqrt{1-x_W}\left[\frac{3\kappa_\phi^2}{4}+\kappa_\Phi^2\left(1-x_W+\frac{3}{4}x_W^2\right)\left(1-\frac{6{\xi}m_r^2}{m_r^2-m_h^2}\right)^2\right.\notag\\
&\qquad\qquad\qquad\left.-\kappa_\phi\kappa_\Phi\left(1+\frac{1}{2}x_W\right)\left(1-\frac{6{\xi}m_r^2}{m_r^2-m_h^2}\right)\right],\label{eq:WW-width}\\[0.5cm]
\Gamma(r{\rightarrow}ZZ)&=\frac{1}{2}\Gamma(r{\rightarrow}W^+W^-),\\[0.5cm]
\Gamma(r{\rightarrow}hh)&=\frac{m_r^3}{32{\pi}M^2}\sqrt{1-x_h}\left[\frac{\kappa_\phi^2}{4}+\kappa_\Phi^2\left(1+x_h+\frac{1}{4}x_h^2-12\xi-6{\xi}x_h+36\xi^2\right.\right.\notag\\
&\qquad\qquad\qquad\left.\left.-\frac{36{\xi}m_h^2}{m_r^2-m_h^2}-\frac{72{\xi}m_h^4}{m_r^2(m_r^2-m_h^2)}+\frac{216\xi^2m_h^2}{m_r^2-m_h^2}+\frac{324\xi^2m_h^4}{(m_r^2-m_h^2)^2}\right)\right.\notag\\
&\qquad\qquad\qquad\left.-\kappa_\phi\kappa_\Phi\left(1+\frac{1}{2}x_h-6\xi-\frac{18{\xi}m_h^2}{m_r^2-m_h^2}\right)\right],\\[0.5cm]
\Gamma(r\rightarrow\bar{f}f)&=\frac{N_c\kappa_\Phi^2m_f^2m_r}{8{\pi}M^2}\left(1-x_f\right)^{3/2}\left(1-\frac{6{\xi}m_r^2}{m_r^2-m_h^2}\right)^2,\\[0.5cm]
\Gamma(r\rightarrow\gamma\gamma)&=\frac{m_r^3}{64{\pi}M^2}\bigg\vert\kappa_\phi+\frac{\alpha_{EM}\kappa_\Phi}{2\pi}\bigg[b_2+b_Y\notag\\
&\qquad\qquad\qquad-\left(1-\frac{6{\xi}m_r^2}{m_r^2-m_h^2}\right)\left(2+3x_W+3x_W(2-x_W)f(x_W)\right)\notag\\
&\qquad\qquad\qquad+\frac{8}{3}\left(1-\frac{6{\xi}m_r^2}{m_r^2-m_h^2}\right)x_t\left(1+(1-x_t)f(x_t)\right)\bigg]\bigg\vert^2,\label{eq:gamma-width}\\[0.5cm]
\Gamma(r{\rightarrow}gg)&=\frac{m_r^3}{8{\pi}M^2}\bigg\vert\kappa_\phi+\frac{\alpha_s\kappa_\Phi}{2\pi}\left[b_{QCD}+\left(1-\frac{6{\xi}m_r^2}{m_r^2-m_h^2}\right)x_t\left(1+(1-x_t)f(x_t)\right)\right]\bigg\vert^2,\label{eq:gg-width}
\end{align}
where
\begin{equation}
f(x)=\left\{\begin{aligned}
&\left[\sin^{-1}\left(\frac{1}{\sqrt{x}}\right)\right]^2,& \text{$x\geq1$},\\
&-\frac{1}{4}\left[\ln\frac{1+\sqrt{1-x}}{1-\sqrt{1-x}}-i\pi\right]^2,& \text{$x<1$},
\end{aligned}
\right.
\end{equation}
and we have defined $x_i=4m_i^2/m_r^2,\ (i=W,Z,h,f)$ and $N_c=3(1)$ for quarks (leptons). When calculating the partial decay widths to $gg$ and $\gamma\gamma$, we have also included the one-loop contributions of the top quark and the W boson, which become important when the mixing is strong. We have neglected one-loop corrections proportional to $\kappa_\phi$. Finally, in the limit $\xi=0$ and $\kappa_\phi\rightarrow0$ our results reduce to those for the RS case given in \cite{Bae:2000pk,Cheung:2000rw}.

\subsection{Branching fractions}
\begin{figure}[H]
\begin{minipage}[b]{0.5\linewidth}
\begin{center}
\includegraphics[height=5cm]{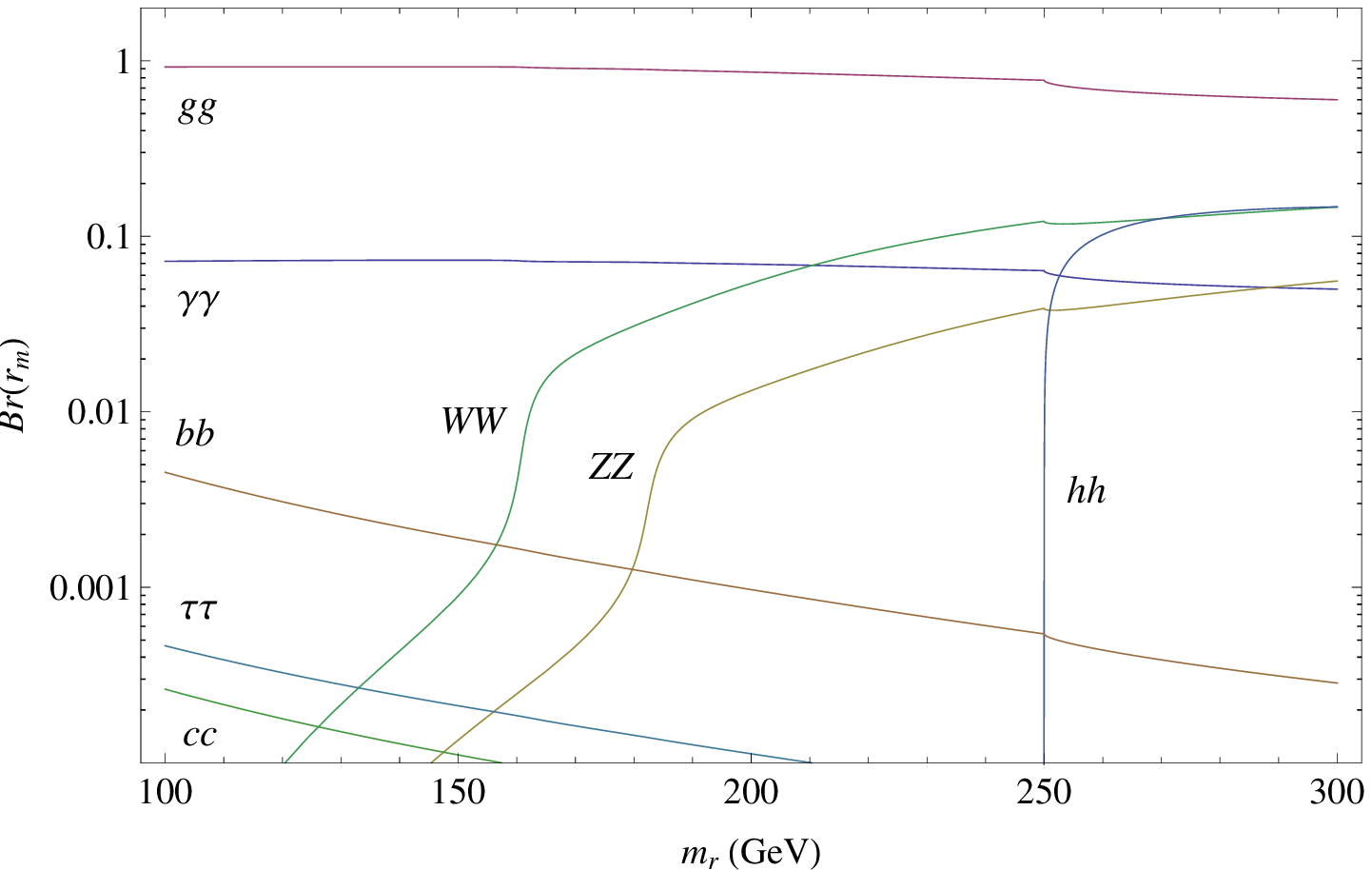}
\end{center}
\end{minipage}
\begin{minipage}[b]{0.5\linewidth}
\begin{center}
\includegraphics[height=5cm]{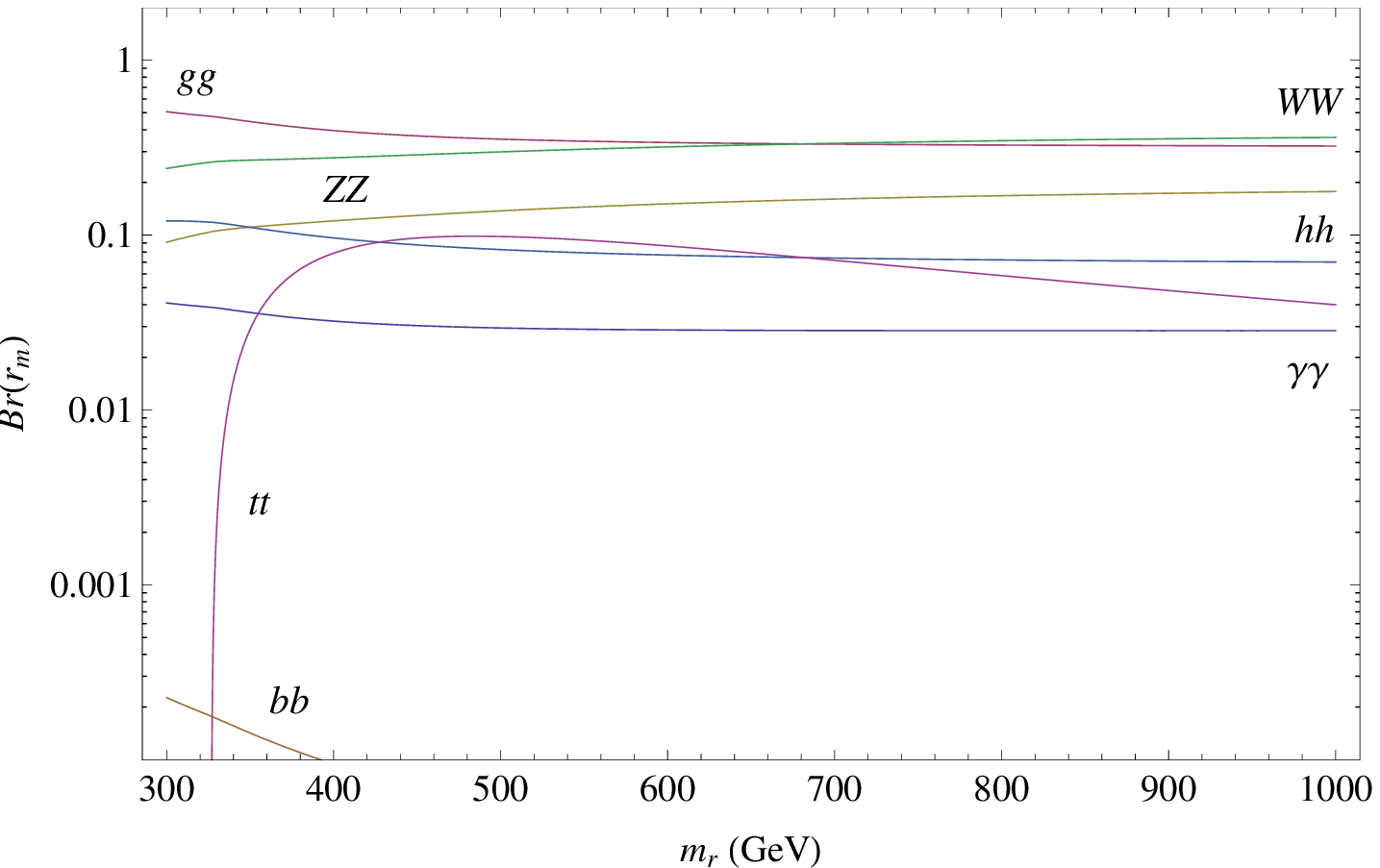}
\end{center}
\end{minipage}
\caption{Branching fractions of $r_m$ as a function of its mass with $\epsilon_{vis}=3$, $M=4\ $TeV, $\xi=0$ and $m_h=125$ GeV. The left and right panels are the same but cover a different range in mass.}
\label{fig:no-mix}
\end{figure}

In this section we use our previous results to look at the branching fractions of the various radion decay modes. We begin by considering the case where there is no mixing between the radion and the Higgs ($\xi=0$). Figure~\ref{fig:no-mix} shows the radion branching fractions as a function of its mass where we have taken $\epsilon_{vis}=3$ and $M=4\ $TeV. We have chosen a value for the Higgs mass of 125 GeV.

We see that in the low mass range the branching fractions are dominated by the $gg$ and $\gamma\gamma$ channels, while in the high mass range $gg$ and $WW$ are the dominant decay modes, but with significant contributions from several other channels. The large branching fraction to $\gamma\gamma$ is of particular interest and is a direct result of the dilaton coupling to the gauge boson kinetic terms. It also remains significant even at high radion masses.

If we compare the branching fractions for the linear dilaton model shown here with those of the RS model \cite{Bae:2000pk}, we note that the most important difference comes in the branching fraction to $\gamma\gamma$. This is significantly enhanced in the linear dilaton  model, since the dilaton coupling is much stronger than the scale anomaly and one-loop contributions, which are proportional to $\alpha_{EM}$. We conclude that the large branching fraction to photons seen in figure~\ref{fig:no-mix} is therefore an important distinguishing feature of this model. 

\begin{figure}[H]
\begin{minipage}[b]{0.5\linewidth}
\begin{center}
\includegraphics[height=5cm]{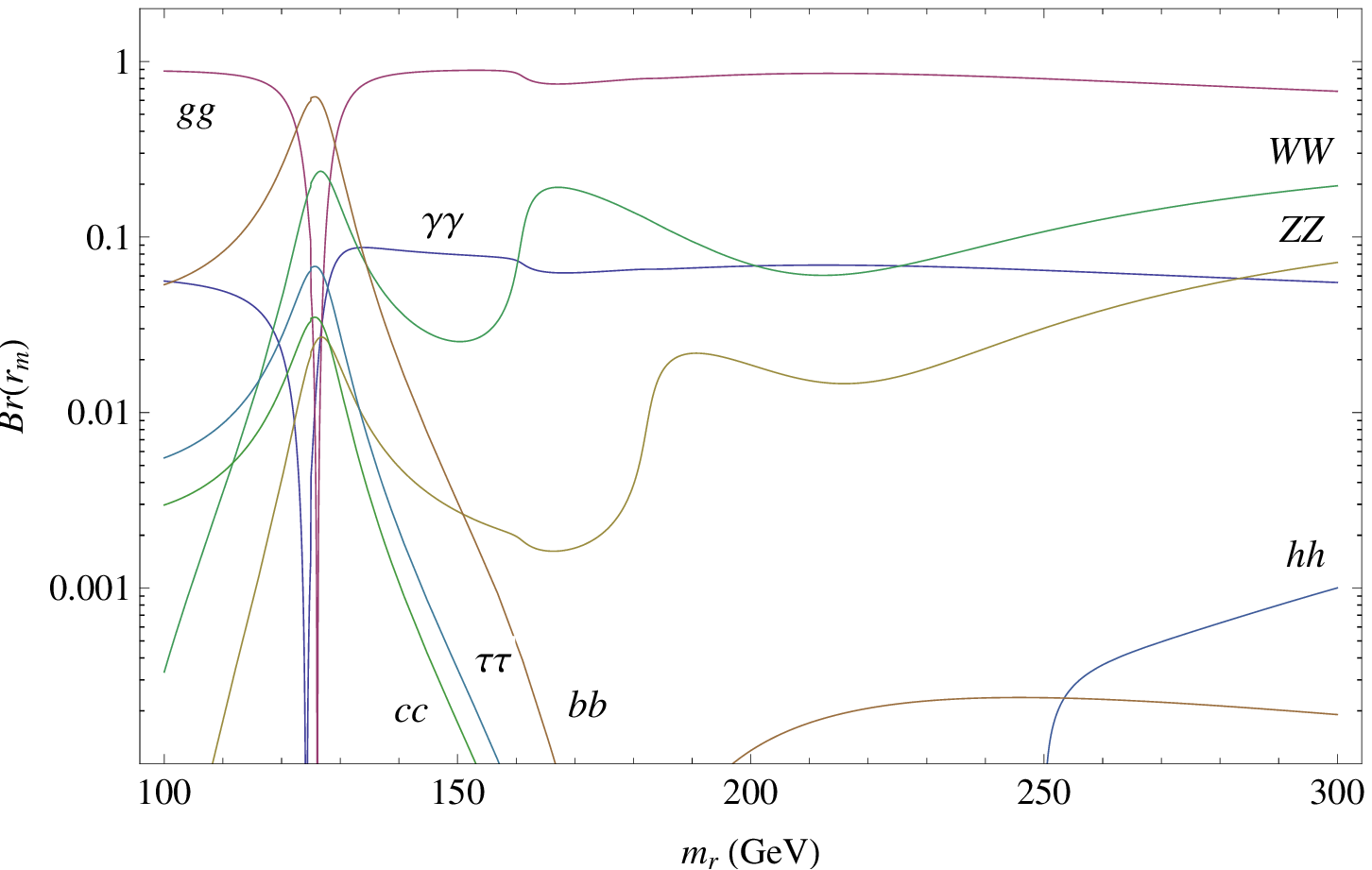}
\end{center}
\end{minipage}
\begin{minipage}[b]{0.5\linewidth}
\begin{center}
\includegraphics[height=5cm]{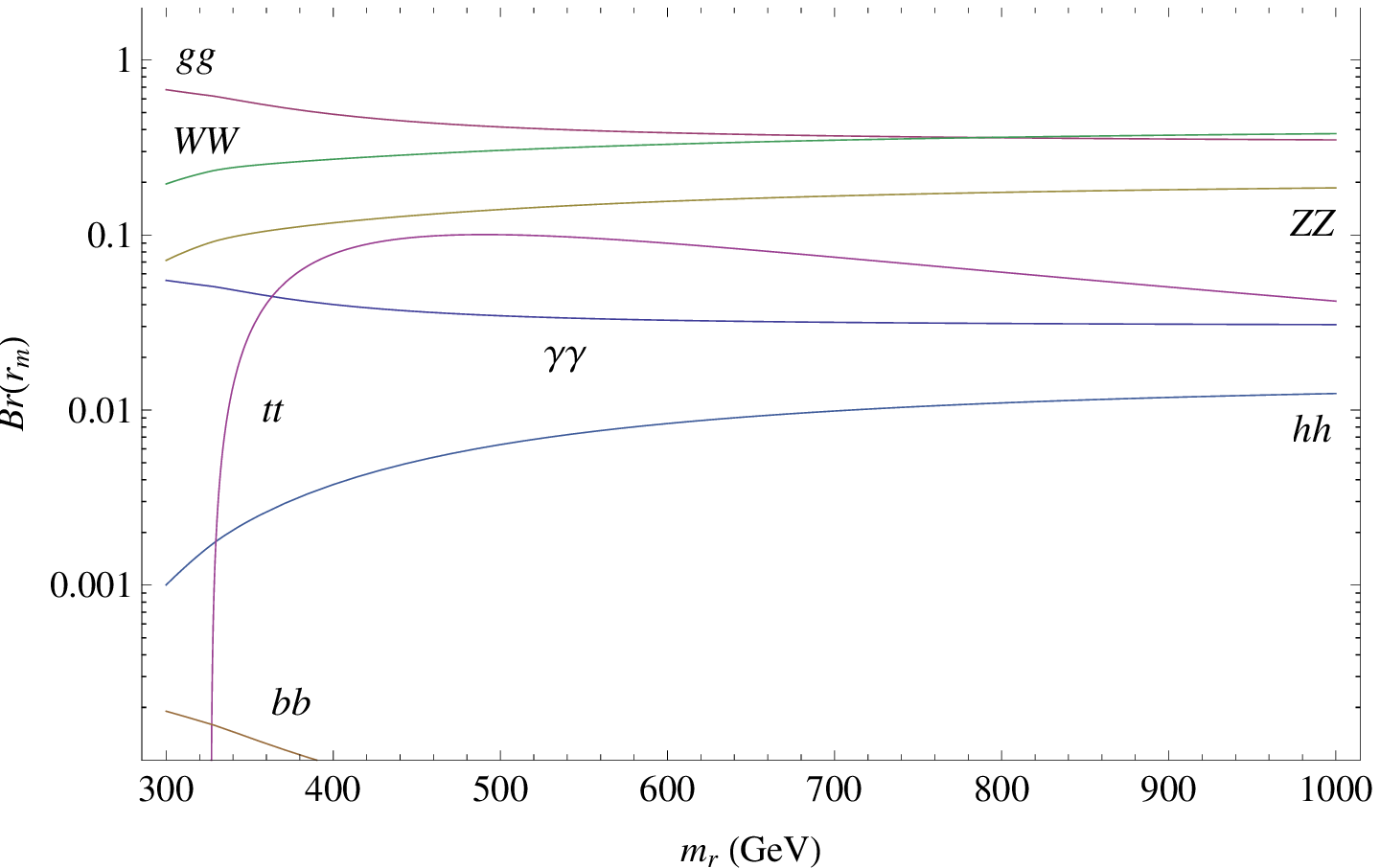}
\end{center}
\end{minipage}
\caption{Branching fractions of $r_m$ as a function of its mass with $\epsilon_{vis}=3$, $M=4\ $TeV, $\xi=1/6$ and $m_h=125$ GeV. The left and right panels are the same but cover a different range in mass.}
\label{fig:mix}
\end{figure}

Next we look at the case where there is a mixing between the radion and Higgs. We shall take $\xi=1/6$, which corresponds to the conformal limit when we also take $m_h$=0. The radion branching fractions are then shown in figure~\ref{fig:mix}. We observe that at high mass, the situation is similar to the zero mixing case in figure~\ref{fig:no-mix}, except that the $hh$ decay mode is now suppressed as a result of the mixing. In the low mass region we observe an interesting drop in the branching fractions to $gg$ and $\gamma\gamma$ when the radion mass is near to that of the Higgs. This is the result of a cancellation in the partial widths \eqref{eq:gamma-width}, \eqref{eq:gg-width} due to the strong mixing in this region.

\subsection{Production}
\begin{figure}[H]
\begin{center}
\includegraphics[height=7cm]{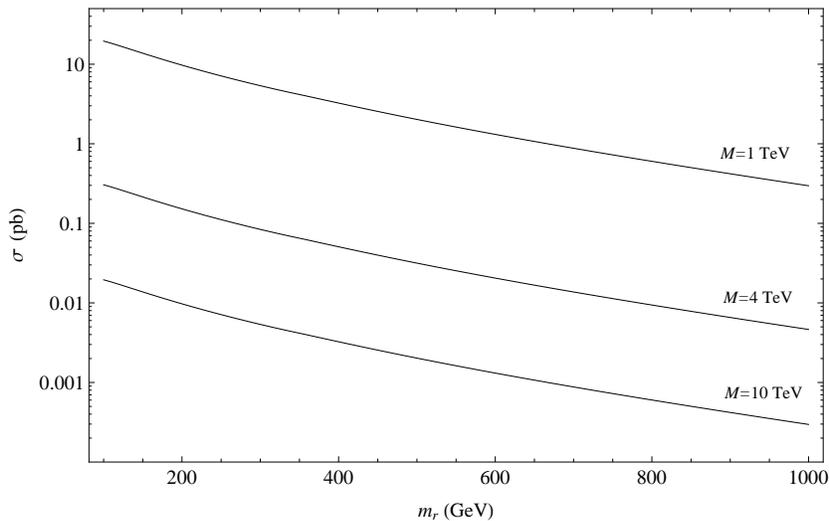}
\caption{Gluon-gluon fusion production cross section for the radion at the LHC with a centre of mass energy of 8 TeV. We have taken $\epsilon_{vis}=3$ and $\xi=0$.}
\label{fig:production}
\end{center}
\end{figure}

In this section we briefly discuss the production of the radion at colliders. At hadronic colliders, such as the LHC, radion production will be dominated by gluon-gluon fusion. The production cross section for this process at a hadron collider with centre of mass energy $\sqrt{s}$ is given by
\begin{equation}
\sigma(pp{\rightarrow}r)=\int^{1}_{m_r^2/s}\frac{dx}{x}g(x,m_r)g\left(\frac{m_r^2}{sx},m_r\right)\frac{m_r^2}{s}\hat{\sigma}(gg{\rightarrow}r)
\end{equation}
where $g(x,q)$ is the gluon parton distribution function at momentum fraction $x$ and renormalisation scale $q$, and $\hat{\sigma}(gg{\rightarrow}r)$ is the gluon-gluon fusion subprocess cross section, which is given by
\begin{equation}\label{eq:production}
\hat{\sigma}(gg{\rightarrow}r)=\frac{\pi}{64M^2}\bigg\vert\kappa_\phi+\frac{\alpha_s}{2\pi}\kappa_\Phi\bigg(b_{QCD}+x_t\big(1+(1-x_t)f(x_t)\big)\bigg)\bigg\vert^2.
\end{equation}
The gluon-gluon fusion production cross section at the LHC for a centre of mass energy of 8 TeV is shown in figure~\ref{fig:production} for several values of the fundamental sale, $M$. The parameterisation of CTEQ5L \cite{Lai:1999wy} has been used for the parton distribution functions. We find that even for relatively large values of $M$ the radion production cross section remains significant. This suggests good prospects for searches at the LHC, as we shall see in the following section.

\subsection{Constraints on parameter space}
Similarities between the radion and the Higgs boson enable us to place constraints on the parameters of the model using the results from Higgs searches currently underway at the LHC. Similar studies have recently been performed for the RS radion in \cite{Davoudiasl:2010fb,Barger:2011qn,Barger:2011nu,Coleppa:2011zx,deSandes:2011zs,Cheung:2011nv}. We begin by using a Breit-Wigner narrow width approximation to relate the radion and Higgs cross sections according to
\begin{equation} \label{eq:DR}
\frac{\sigma(pp{\rightarrow}r{\rightarrow}X)}{\sigma_{SM}(pp{\rightarrow}h{\rightarrow}X)}=\frac{\Gamma(r{\rightarrow}gg)}{\Gamma(h{\rightarrow}gg)}\frac{Br(r{\rightarrow}X)}{Br(h{\rightarrow}X)}.
\end{equation}

\begin{figure}[H]
\begin{center}
\includegraphics[height=7cm]{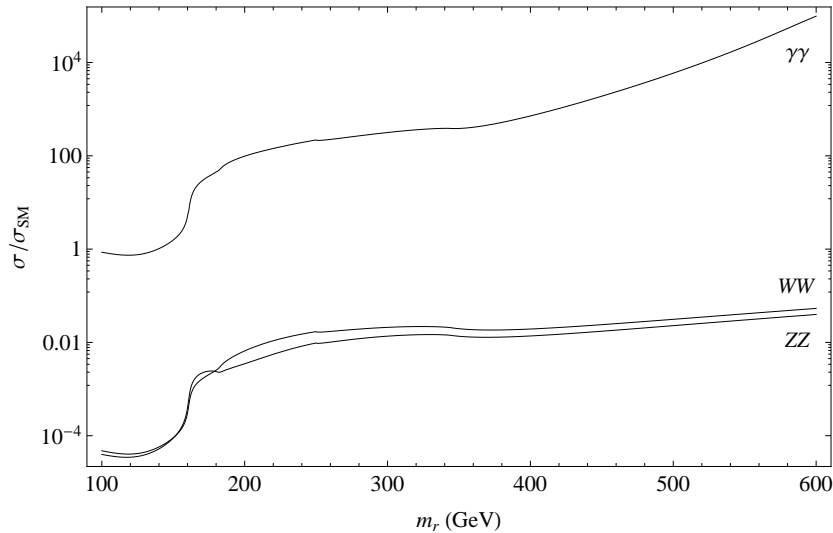}
\caption{$\sigma/\sigma_{SM}$ as a function of the radion mass. We have taken $\epsilon_{vis}=3$, $M=4\ $TeV and $\xi=0$.}
\label{fig:DR}
\end{center}
\end{figure}

This ratio is plotted as a function of the radion mass for the $WW$, $ZZ$ and $\gamma\gamma$ decay channels in figure~\ref{fig:DR}. We observe that the strongest constraints on the model will be obtained using the $\gamma\gamma$ decay channel, which gives a value of $\sigma/\sigma_{SM}$ of order one, due to the large branching fraction of the radion to this mode.

\begin{figure}[H]
\begin{center}
\includegraphics[height=7cm]{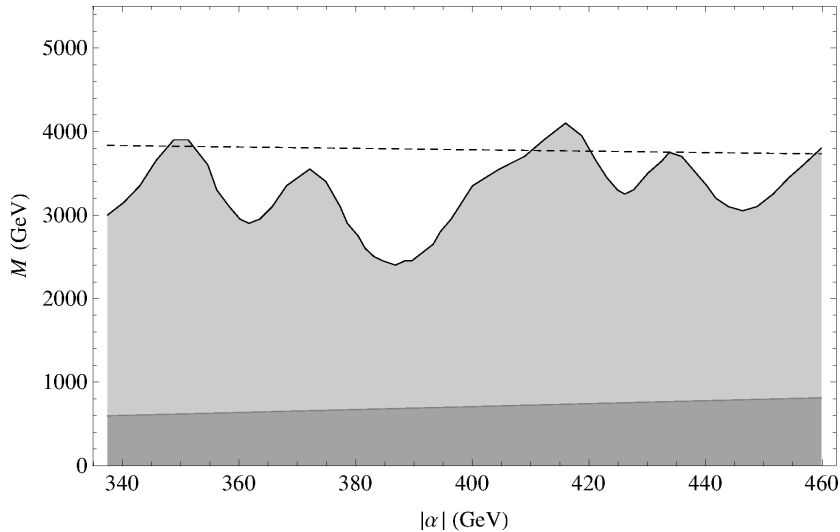}
\caption{Constraints on the model parameters, $\alpha$ and $M$, using ATLAS results in the $h\rightarrow\gamma\gamma$ channel. The lighter shaded region is excluded by the data and the darker region is the theoretically excluded region where $M^2<\frac{28}{9}\alpha^2$. The dashed line shows the constraints from ADD total cross section searches at the LHC~\cite{Baryakhtar:2012wj}. We have taken $\epsilon_{vis}=3$.}
\label{fig:Malimit}
\end{center}
\end{figure}

We use the latest results from the ATLAS experiment in the $h\rightarrow\gamma\gamma$ channel with 4.9 fb$^{-1}$ integrated luminosity \cite{Collaboration:2012sk}, where the CL$_s$ method has been used to place bounds on $\sigma/\sigma_{SM}$ at the 95\% confidence level. Comparing these results to \eqref{eq:DR} we can place constraints on the parameters of the model, namely the fundamental scale, $M$, and the curvature scale, $\alpha$.  These constraints are shown in figure~\ref{fig:Malimit}, where the lighter shaded region is excluded by the data, and the darker region is the theoretically excluded region where the 5D curvature, $\frac{28}{9}\alpha^2$ is larger than $M^2$. The dashed line shows the constraints from ADD total cross section searches at the LHC~\cite{Baryakhtar:2012wj}.

We obtain a limit of $M\gtrsim3\ $TeV for values of $\vert\alpha\vert$ in the range 340-460 GeV. Note however that these limits will depend on the value taken for $\epsilon_{vis}$, which is not constrained. For larger values of $\epsilon_{vis}$ the limits become stronger, while for $\epsilon_{vis}\lesssim3$ the ADD total cross section searches provide the best constraints on $\alpha$ and $M$. Using the $WW{\rightarrow}l{\nu}l{\nu}$ and $ZZ{\rightarrow}4l$ channels, constraints can also be placed on the parameter space in the higher mass regions. However, due to the smaller value of $\sigma/\sigma_{SM}$ for these channels, the limits currently lie below the theoretical exclusion.

\section{Conclusion}
Motivated by Little String Theory, the 5D linear dilaton model provides a distinctive solution to the hierarchy problem. It is characterised by a graviton KK mass spectrum which consists of a $\sim$ TeV scale mass gap,
followed by a near continuum of resonances spaced $\sim$ 30 GeV apart. The scalar sector
of the model plays a crucial role, since it contains a scalar field, the dilaton, which can be used to stabilise the extra dimension. Just like in the Goldberger-Wise mechanism,
the dilaton acquires a non-zero vacuum expectation value on the branes after imposing appropriate boundary potentials. A large volume is then stabilised with an interbrane proper distance of $y_c\sim10^{10}\vert\alpha\vert^{-1}$. The correct hierarchy is achieved without the need for any extreme fine tuning. 

The coupled scalar fluctuations about the metric and dilaton fields were then considered. We solved the equations of motion for the system and obtained the wavefunctions and masses of the modes. It was found that the radion and lowest KK mode both obtain masses of order the curvature scale, but can be parametrically lighter depending on the boundary mass parameters. The KK modes were then closely spaced and essentially formed a near continuum of modes of order the curvature scale and above.

This model introduces a direct coupling in the boundary action between the dilaton and the Standard Model fields. It was shown that this results in a coupling between the radion and SM Lagrangian, in addition to the standard coupling to the trace of the energy-momentum tensor. The strength of these couplings was calculated, with a typical strength of order $(10\ \mathrm{TeV})^{-1}$, and a suppression in the coupling to the SM Lagrangian by a factor inversely proportional to the boundary mass term. Additionally, the couplings of the KK modes were suppressed relative to the radion, so that we would initially expect to observe only a single mode in experiments, despite the closely spaced mass spectrum.

We also considered the situation where the Higgs couples non-minimally to gravity and included an additional Higgs-curvature interaction term in the 4D effective action, resulting in a kinetic mixing between the radion and the Higgs. However, we found that at leading order the dilaton did not contribute to the mixing and our results reduced to those for the RS case. The mixing then had virtually no effect on the Higgs phenomenology, except in the case when the difference between the radion and Higgs masses was small. The contribution of the KK modes to the mixing was also investigated. However, we found that despite their close mass spacings, the cumulative effect of the KK modes was still too small to significantly effect the Higgs phenomenology.

The decay widths and branching fractions of the radion were calculated in section~\ref{sec:decays} and found to be substantially different to the RS radion. This was largely due to the additional coupling between the radion and the gauge boson kinetic terms. In particular this led to a significantly increased branching fraction to $\gamma\gamma$, which provides an interesting distinguishing feature of this model. 

The production of the radion via gluon-gluon fusion at the LHC was also discussed and found to be significant even for relatively large values of $M$. Finally, we demonstrated the potential of current Higgs searches at the LHC to constrain the parameter space of this model. Using the latest ATLAS results in the $h\rightarrow\gamma\gamma$ channel we obtained a limit of $M\gtrsim3\ $TeV for values of $\vert\alpha\vert$ in the range 340-460 GeV when $\mu_{vis}=\frac{1}{3}\vert\alpha\vert$.

\subsection*{Acknowledgments}
We thank Marco Peloso and Nicholas Setzer for helpful discussions as well as Anibal Medina for helpful discussions and comments on the manuscript. We are grateful to Hooman Davoudiasl for pointing out an error in the radion-gauge boson Feynman rules in the first version of this paper. This work was supported in part by the Australian Research Council. TG also thanks Savas Dimopoulos and the SITP at Stanford for support and hospitality during the completion of this work.

\appendix
\section{Wavefunction Normalisation}

In order to calculate the couplings to the Standard Model fields we need to ensure that the $Q_n$ modes are canonically normalised, which corresponds to correctly setting the normalisation constant $N_n$. However, the equations of motion are only linear in the perturbations and do not contain sufficient information to correctly normalise the modes. It therefore becomes necessary to diagonalise the action \eqref{eq:action} to second order in the perturbations. This analysis has been done for the general case in \cite{Kofman:2004tk}. Using their result, we obtain the following free action for the 4D physical modes of the system,
\begin{equation}
S_n^{free}=C_n{\int}d^4x\, Q_n\left[\Box-m_n^2\right]Q_n,
\end{equation}
\begin{equation}\label{eq:Cn}
C_n=\frac{27M^3}{2\alpha^2}\int_0^{r_c}dz\, e^{-{\alpha}z}\left[\Phi_n'^2-\frac{4\alpha}{3}\Phi_n'\Phi_n+\frac{2\alpha^2}{3}\Phi_n^2\right]\equiv\frac{1}{2}.
\end{equation}

In addition to correctly normalising the modes, diagonalising the action to second order in the perturbations allows us to determine the precise linear combination of the fields $\Phi$, $\delta\phi$, which corresponds to the 5D dynamical variable in the bulk. 
\begin{equation}
\chi=2\sqrt{3} e^{-\frac{1}{2}{\alpha}z}\left(\frac{\Phi}{2}-\frac{\delta\phi}{3}\right),
\end{equation}
where $\chi$ is the dynamical bulk field and obeys the following equation of motion
\begin{equation}
\left[\Box+\frac{d^2}{dz^2}-\frac{\alpha^2}{4}\right]\chi=0.
\end{equation}

Here we provide the complete normalised wavefunctions for the radion and KK modes. The wavefunctions take the form of \eqref{eq:wavefunction}, where the normalisation constant $N_n$ is given by
\begin{multline}
N_n=\frac{4\sqrt{\beta_n}}{3M^{3/2}}\bigg[4\beta_n^2\alpha\epsilon+\alpha^3(1+\epsilon)\bigg]\bigg[6{\beta_n}r_c(4\beta_n^2+\alpha^2)(16\beta_n^4\epsilon^2+4\beta_n^2\alpha^2(9+2\epsilon+2\epsilon^2)+\alpha^4(1+\epsilon)^2)\\
-8\beta_n\alpha\bigg(16\beta_n^4\epsilon(9+\epsilon)+8\beta_n^2\alpha^2\epsilon(1+\epsilon)+\alpha^4(-8-7\epsilon+\epsilon^2)\bigg)\sin({\beta_n}r_c)^2\\
+3\bigg(64\beta_n^6\epsilon^2+16\beta_n^4\alpha^2(-9-2\epsilon+\epsilon^2)-4\beta_n^2\alpha^4(-6+4\epsilon+\epsilon^2)-\alpha^6(1+\epsilon)^2\bigg)\sin(2{\beta_n}r_c)\bigg]^{-1/2},
\end{multline}
where we have defined $\epsilon\equiv\epsilon_{vis}$ for simplicity and $\beta_n$ is defined near \eqref{eq:wavefunction} and can depend on both $\epsilon_{vis}$ and $\epsilon_{hid}$.

\section{Higgs-radion KK mixing}
\label{sec:KKmixing}
We consider the effect of the Higgs mixing with the closely spaced radion KK modes.
The Lagrangian that needs to be diagonalised is now given by
\begin{multline}
\mathcal{L}=-\frac{1}{2}h{\Box}h-\frac{1}{2}m_h^2h^2+\sum_n\left(-\frac{1}{2}Q_n{\Box}Q_n-\frac{1}{2}m_n^2Q_n^2+\frac{6{\xi}\kappa_{\Phi,n}v}{M}h{\Box}Q_n\right)\\
-\frac{3{\xi}v^2}{M^2}\sum_m\sum_n\kappa_{\Phi,m}\left(\kappa_{\Phi,n}-\kappa_{\phi,n}\right)Q_m{\Box}Q_n.
\end{multline}
We see that in addition to the Higgs-KK mixing, the last term also introduces a kinetic mixing between the KK modes. However, this term is suppressed by an additional factor of $1/M$. We shall therefore work to leading order in $1/M$, which allows us to neglect the mixing between the KK modes and focus instead on the mixing with the Higgs. The analysis can now proceed as for the single radion case and we diagonalise the kinetic terms via the transformation
\begin{equation}
h=h'+\frac{6{\xi}v}{M}\sum_n\kappa_{\Phi,n}Q_n.
\end{equation}
The Lagrangian now takes the form
\begin{equation}
\mathcal{L}=-\frac{1}{2}h'{\Box}h'-\frac{1}{2}\sum_nQ_n{\Box}Q_n-\frac{1}{2}f^im^{ij}f^j,
\end{equation}
where we have defined the vector
\begin{equation}
f^i=\left(
\begin{array}{c}
h'\\
r\\
Q_1\\
\vdots\\
\end{array}
\right),
\end{equation}
and the mass matrix is given by
\begin{equation}
m^{ij}=m_h^2\left(
\begin{matrix}
1 & \frac{6{\xi}\kappa_{\Phi,r}v}{M} & \frac{6{\xi}\kappa_{\Phi,1}v}{M} & \hdots\\[0.5em]
\frac{6{\xi}\kappa_{\Phi,r}v}{M} & \frac{m_r^2}{m_h^2} & & \text{\large 0}\\[0.5em]
\frac{6{\xi}\kappa_{\Phi,1}v}{M} & & \frac{m_1^2}{m_h^2}\\[0.5em]
\vdots & \text{\large 0} & & \ddots\\[0.5em]
\end{matrix}
\right).
\end{equation}

It is now straightforward to diagonalise the mass matrix numerically. We find that the mixing with the Higgs remains small and there is one mass eigenstate which is largely composed of the Higgs gauge eigenstate. So despite their relatively close mass spacing, the cumulative effect of the KK modes is still too small to have a significant effect on the Higgs phenomenology. There is however an exception when one of the KK modes has a mass very close to the mass of the Higgs, in which case the mixing between those two states becomes significant. However the mixing with the other KK modes reduces this effect compared to the radion only case discussed previously.

\section{Feynman Rules}
\label{sec:feynrules}

\begin{minipage}[b]{0.5\linewidth} 
\begin{figure}[H]
\includegraphics[height=4cm]{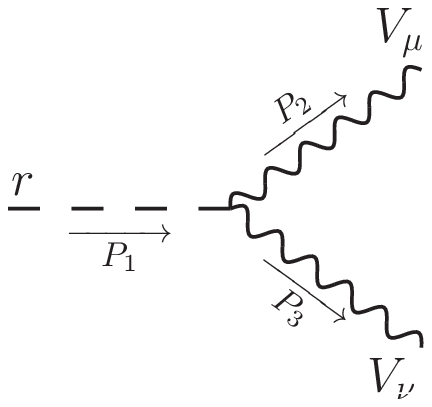}
\end{figure}
\end{minipage}
\begin{minipage}[b]{0.5\linewidth}
\begin{multline}
\frac{ib_1\kappa_\phi}{M}\left(P_2{\cdot}P_3\eta^{\mu\nu}-P_2^{\nu}P_3^{\mu}\right)\\
+2im_V^2\left(\frac{b_1}{M}\left(\frac{\kappa_\phi}{2}-\kappa_\Phi\right)+\frac{a_1}{v}\right)\eta^{\mu\nu}
\end{multline}
\\[0.05cm]
\end{minipage}
\\[2cm]  
\begin{minipage}[b]{0.5\linewidth} 
\begin{figure}[H]
\includegraphics[height=4cm]{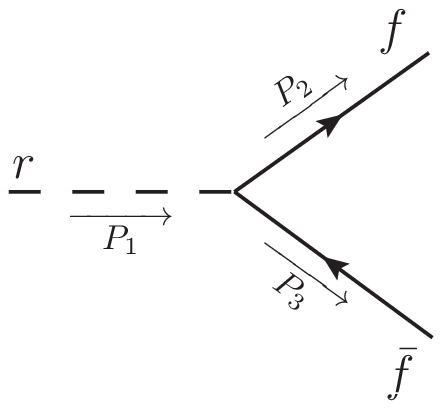}
\end{figure}
\end{minipage}
\begin{minipage}[b]{0.5\linewidth}
\begin{multline}
\frac{ib_1}{M}\left(\kappa_\phi-3\kappa_\Phi\right)\displaystyle{\not}P_2-\frac{3ib_1\kappa_\Phi}{2M}\displaystyle{\not}P_1\\
+im_\Psi\left(\frac{b_1}{M}\left(4\kappa_\Phi-\kappa_\phi\right)-\frac{a_1}{v}\right)
\end{multline}
\\
\begin{equation}
im_\Psi\left(\frac{b_1\kappa_\Phi}{M}-\frac{a_1}{v}\right) \qquad\text{(on-shell)}
\end{equation}
\end{minipage} 
\\[2cm]
\begin{minipage}[b]{0.5\linewidth} 
\begin{figure}[H]
\includegraphics[height=4cm]{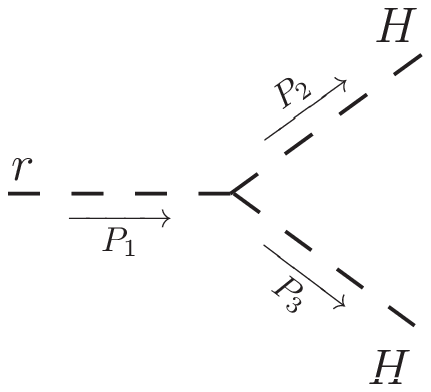}
\end{figure}
\end{minipage}
\begin{minipage}[b]{0.5\linewidth}
\begin{multline}
-\frac{2ia_0^2b_1}{M}\left(\frac{\kappa_\phi}{2}+\kappa_\Phi(6\xi-1)\right)P_2{\cdot}P_3\\
-\frac{6i{\xi}a_0^2b_1}{M}\kappa_\Phi\left(P_2{\cdot}P_2+P_3{\cdot}P_3\right)\\
+2ia_0^2m_h^2\left(\frac{b_1}{M}\left(2\kappa_\Phi-\frac{\kappa_\phi}{2}\right)-\frac{3a_1}{2v}\right)
\end{multline}
\end{minipage} 
\\[2cm]
\begin{minipage}[b]{0.5\linewidth} 
\begin{figure}[H]
\includegraphics[height=4cm]{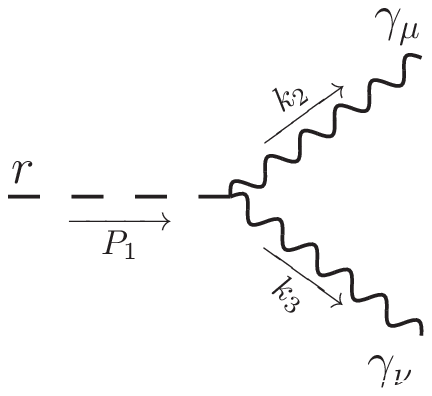}
\end{figure}
\end{minipage}
\begin{minipage}[b]{0.5\linewidth}
\begin{equation}
\frac{ib_1}{M}\left(\kappa_\phi+\frac{\alpha_{EM}}{2\pi}\kappa_\Phi(b_2+b_Y)\right)\left(k_2{\cdot}k_3\eta^{\mu\nu}-k_2^{\nu}k_3^\mu\right)
\end{equation}
\\[0.5cm]
\end{minipage} 
\\[2cm]
\begin{minipage}[b]{0.5\linewidth} 
\begin{figure}[H]
\includegraphics[height=4cm]{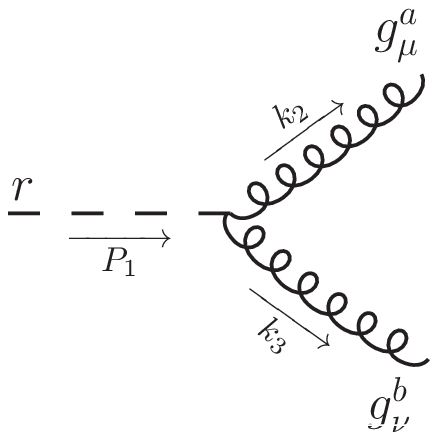}
\end{figure}
\end{minipage}
\begin{minipage}[b]{0.5\linewidth}
\begin{equation}
\frac{ib_1}{M}\left(\kappa_\phi+\frac{\alpha_sb_{QCD}}{2\pi}\kappa_\Phi\right)\left(k_2{\cdot}k_3\eta^{\mu\nu}-k_2^{\nu}k_3^\mu\right)\delta^a_b
\end{equation}
\\[0.15cm]
\end{minipage}

\newpage
\addcontentsline{toc}{section}{References}
\bibliographystyle{biblio_style}
\bibliography{paper_bib}

\end{document}